\documentclass[pra,amsmath,amssymb, aps, reprint, longbibliography]{revtex4-2}
\usepackage[utf8]{inputenc}
\usepackage[T1]{fontenc}

\usepackage{amsthm}
\usepackage{amsmath,bm,bbm}
\usepackage{amssymb}
\usepackage{amsfonts}
\usepackage{braket}
\usepackage{graphicx}
\usepackage{mathtools}
\usepackage[unicode=true, breaklinks=false, pdfborder={0 0 1}, backref=false, colorlinks=true, linkcolor=blue, citecolor=blue, urlcolor=blue]{hyperref}

\usepackage{dsfont}

\usepackage{graphicx} %cropping

\usepackage[normalem]{ulem} %crossing out

\hypersetup{citecolor=magenta}
\hypersetup{colorlinks=true}
\hypersetup{linkcolor=blue}
\hypersetup{urlcolor=blue}

\DeclareMathOperator\tr{tr}

\newcommand{\la}{\langle}
\newcommand{\ra}{\rangle}

\newcommand{\diff}{\mathrm{d}}
\newcommand{\e}{\mathrm{e}}
\renewcommand{\vec}[1]{\boldsymbol{#1}}
\newcommand{\Op}[1]{\hat{#1}}

\newcommand{\ve}{\vec{e}}

\newcommand{\oH}{\Op{H}}

\newcommand{\oL}{\Op{L}}

\newcommand{\oU}{\Op{U}}
\newcommand{\oW}{\Op{W}}
\newcommand{\oV}{\Op{V}}

\newcommand{\ogamma}{\Op{\gamma}}

\newcommand{\Boltz}{k_{\mathrm{B}}}
\newcommand{\Gibbs}{\hat{\gamma}}
\newcommand{\oalpha}{ \hat{\alpha}}

\newcommand{\var}{\text{var}}

\usepackage{xcolor}
\definecolor{darkred}{rgb}{0.7 0.0 0.0}
\newcommand\mg[1]{{#1}}

\begin{document}
\title{Otto cycles with a quantum planar rotor}

\author{Michael Gaida}
\email{michael.gaida@student.uni-siegen.de}
\affiliation{Naturwissenschaftlich--Technische Fakult\"{a}t, Universit\"{a}t Siegen, 
Walter-Flex-Stra{\ss}e 3, 57068 Siegen, Germany}
\author{Stefan Nimmrichter}
\email{stefan.nimmrichter@uni-siegen.de}
\affiliation{Naturwissenschaftlich--Technische Fakult\"{a}t, Universit\"{a}t Siegen, 
Walter-Flex-Stra{\ss}e 3, 57068 Siegen, Germany}

\begin{abstract}
We present two realizations of an Otto cycle with a quantum planar rotor as the working medium controlled by means of external fields. By comparing the quantum and the classical description of the working medium, we single out genuine quantum effects with regards to the performance and the engine and refrigerator modes of the Otto cycle. The first example is a rotating electric dipole subjected to a controlled electric field, equivalent to a quantum pendulum. Here we find a systematic disadvantage of the quantum rotor compared to its classical counterpart. 
In contrast, a genuine quantum advantage can be observed with a charged rotor generating a magnetic moment that is subjected to a controlled magnetic field. Here, we prove that the classical rotor is inoperable as a working medium for any choice of parameters, whereas the quantum rotor supports an engine and a refrigerator mode, exploiting the quantum statistics during the cold strokes of the cycle.

\end{abstract}

\maketitle

\section{Introduction}

With the thermodynamic interpretation of the three level maser in 1959  \cite{scovil_three-level_1959}, the field of quantum thermodynamics was born. Since then, quantum analogues of classical thermal machine models such as the Carnot cycle \cite{geva1992quantum,bender2002entropy,esposito2010efficiency,dann2020quantum, denzler2021power} and the Otto cycle \cite{feldmann1996heat,scully2002quantum,henrich2007quantum, he2009performance,Kosloff2017,son2021monitoring,piccitto2022ising} have been studied exhaustively. In these machines, the classical macroscopic working medium, usually a gas, is replaced by quantum systems of varying complexity that undergo a controlled cycle of strokes, which alternate between the thermal coupling with a hot and a cold reservoir and the modulation of the system Hamiltonian over time, mimicking the (model-extrinsic) motion of a ``piston''. In this work, we focus on the Otto cycle, which can be operated both as an engine that outputs work at the expense of heat from the hot reservoir and as a refrigerator that extracts heat from the cold reservoir at the expense of work.

Studied quantum working media range from small finite-dimensional systems \cite{Thomas2011,Rogerio2019,Peterson2019} to many-body systems \cite{Azimi2014,Yungern2019,Hartmann2020} and infinite-dimensional continuous-variable systems \cite{Kosloff2017, Rezek_2006, Deffner2018}. Experimental proof-of-principle realizations of the Otto cycle were performed with trapped ions \cite{rossnagel2016single,PhysRevLett.123.080602}, nano beams \cite{PhysRevX.7.031044}, nitrogen vacancy centers \cite{PhysRevLett.122.110601}, spins with nuclear magnetic resonance techniques \cite{Peterson2019}, optomechanical systems \cite{PhysRevLett.112.150602}, quantum gases \cite{bouton2021quantum} and proposed for nanomechanical resonators\cite{PhysRevE.95.022135}, circuit QED \cite{karimi2016otto}, and quantum dots \cite{PhysRevE.101.012116}.

Continuous-variable working media naturally lend themselves to the study of quantum effects on the machine performance. They can represent motional degrees of freedom with a classical analogue, e.g.~the position and momentum of a particle, admitting a direct comparison between the classical and the quantum version of the studied machine. Note however that the quantum-classical comparison is often understood as a comparison of  machine models with and without coherence on a given quantum system \cite{uzdin2015equivalence,PhysRevE.99.062102}, provided that the periodic piston modulation affects the energy basis of the working medium. 
While the advantages of quantum machines are often highlighted in specific cases \cite{solfanelli_quantum_2023,sur2023quantum,rolandi2023collective,tajima2021superconducting,kamimura_quantum-enhanced_2022}, the performance of a machine model can in general both improve and deteriorate due to quantization of the working medium. 

One predominantly studied working medium is the harmonic oscillator, due to its mathematically well-understood behaviour. Unfortunately, it was shown that the harmonic oscillator does not have the capacity for genuine quantum effects on the range of parameters at which a standard Otto cycle operates as an engine or refrigerator: A homogeneous scaling of the energy levels with respect to the work parameter $\lambda$ representing the periodic piston modulation of the harmonic frequency, $E_n(\lambda) = \lambda \hbar\omega (n + 1/2)$ implies classical operation modes \cite{gelbwaser-klimovsky_single-atom_2018}. Moreover, the same applies to any Otto cycle in which the cyclic modulation of the Hamiltonian implies a simple proportionality of the energy spectrum, $E_n(\lambda) \propto \lambda^k$.

Here we will consider the quantum planar rotor as a continuous-variable working medium for the quantum Otto cycle. This is in contrast to and a complement of previous theoretical works employing the rotor as an autonomous piston degree of freedom for engines \cite{PhysRevE.95.062131, seah2018work,roulet2018autonomous, puebla2022open, PhysRevE.109.024108}. Angular momentum quantization can lead to genuinely non-classical phenomena in the free evolution of a single planar rotor \cite{stickler2018probing} as well as in the dynamics of coherently interlocked rotors \cite{PhysRevE.99.042202}. Experimental demonstrations of rotor-based machines could be based on molecular rotors, the quantum dynamics of which is nowadays routinely observed and controlled with help of tailored laser pulses \cite{koch2019quantum,Park2015} . Another platform to realize planar rotor analogues and rotor engines is circuit QED \cite{jain1984mutual}, where the Josephson phase plays the role of the rotor angle. 
Finally, state-of-the-art experiments in levitated optomechanics with rigid nanorotors are steadily approaching the quantum regime \cite{Kuhn:17, stickler_quantum_2021,rademacher2022measurement,arita2023cooling,kamba2023nanoscale}.

\begin{figure}
    \centering
 \includegraphics[width=0.45\textwidth]{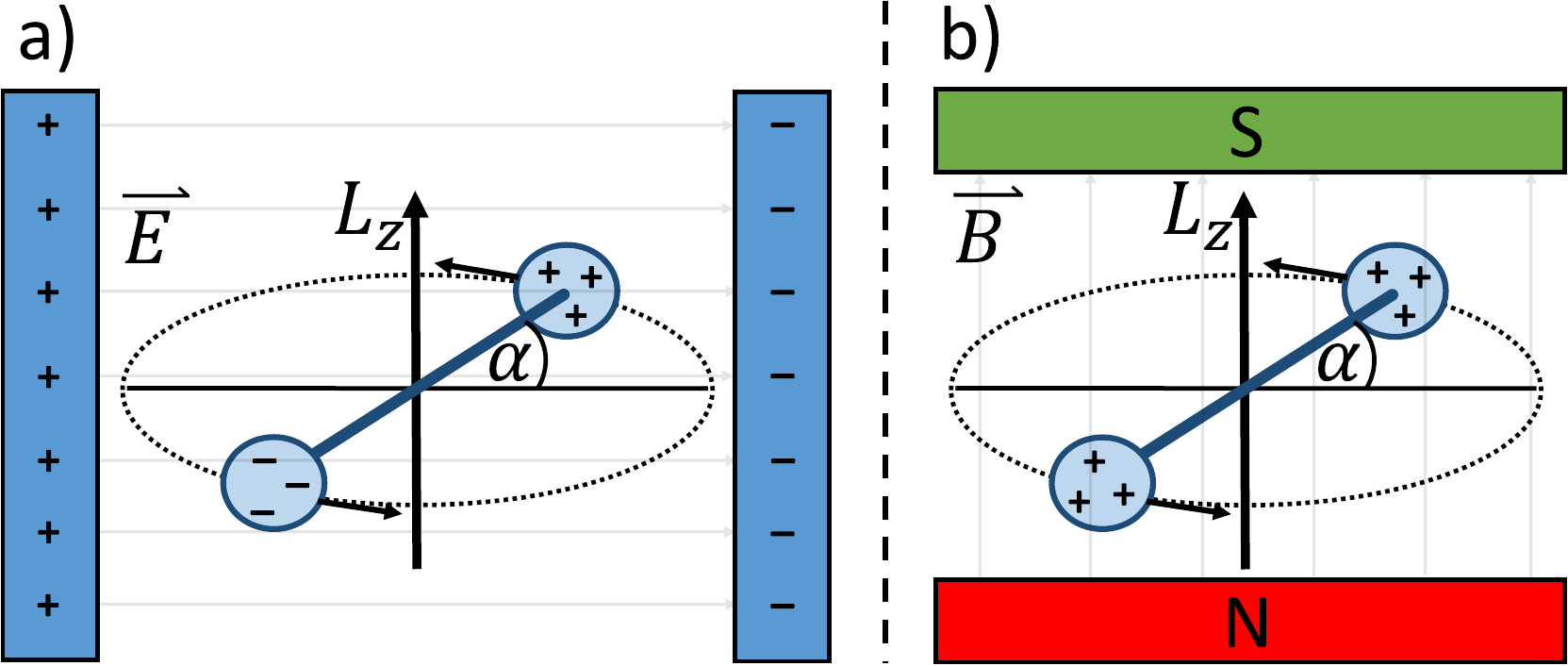}
 \caption{Two physical settings in which a planar rotor acts as the working medium of an Otto cycle: (a) Rotating electric dipole subject to an externally controlled, homogeneous electric field parallel to the rotation plane; (b) Charged rotor generating a magnetic dipole moment subject to a controlled magnetic field perpendicular to the rotation plane.}
    \label{fig_Engines}
\end{figure}

We will formulate the theoretical model of a planar rotor subjected to an externally modulated potential in the four strokes of the paradigmatic Otto cycle (Section \ref{sec_model}) and investigate the classical and quantum predictions for two physically motivated examples, sketched in Fig.~\ref{fig_Engines}: The first one is a rotating electric dipole in the presence of an electric field of alternating strength in the plane of rotation (Section \ref{sec_el}). We show that the so defined Otto cycle, assuming ideal quenches of the electric field and full thermalization in between, always performs worse in the quantum case. Both the operation regimes as an engine or refrigerator and the energy output decrease in comparison to the classical case.
The second example is a dipole in a magnetic field of alternating strength perpendicular to the rotation plane (Section \ref{sec_mag}). Here we show that the classical rotor operates neither as an Otto engine nor as a refrigerator, whereas the quantum rotor does, when the cold stroke of the Otto cycle is operated in the low-excitation regime. This constitutes a genuine advantage enabled by angular momentum quantization. We briefly conclude our study in Section \ref{sec_conclusions}.

\section{Theoretical model} 
\label{sec_model}

A planar rotor is a dynamical degree of freedom described by a single angular coordinate $\alpha \in [0,2\pi)$ and its conjugate angular momentum $L_z$. It represents the phase space of, for instance, a particle on a ring in the $xy$-plane, the phase variable in a superconducting Josephson loop, and the orientation of a rigid rotor on a fixed plane of rotation. 
We will first introduce the notation and theoretical description of a classical and quantum planar rotor, before briefly reviewing the ideal four-stroke Otto cycle.

\subsection{Classical and quantum planar rotor}
\label{sec_rotor}

Classically, the canonical variables $(\alpha, L_z)$ of the planar rotor can be treated in the same manner as the position and momentum of linear motion in one dimension. Physical states are described by phase-space probability densities  $P(\alpha, L_z)$ and valid (time-independent) Hamiltonians by $H(\alpha, L_z)$, with the additional requirement of strict $2 \pi$-periodicity in $\alpha$. We will consider Hamiltonians of the form
\begin{equation}\label{eq:HamCl}
    H = H(\alpha,L_z) = \frac{(L_z - I\omega)^2}{2I} + V(\alpha),
\end{equation}
with a given moment of inertia $I$, a periodic potential, $V(\alpha) = V(\alpha+2\pi)$, and an angular momentum displacement by $I\omega$. The latter can be viewed as a boost with respect to a rotating frame at angular frequency $\omega$. 

The Gibbs state of such a classical planar rotor in thermal equilibrium at temperature $T$ is given by
\begin{equation}\label{eq:classGibbs}
    P(\alpha,L_z) = \frac{1}{Z} \e^{-H(\alpha,L_z)/\Boltz T},
\end{equation}
with the partition function 
\begin{eqnarray}
    Z &=& \int_0^{2\pi}\!\! \diff \alpha \int_{-\infty}^{\infty}\!\!  \diff L_z \, \e^{- H(\alpha,L_z)/\Boltz T} \nonumber \\
    &=& \sqrt{2\pi I \Boltz T} \int_0^{2\pi}\!\! \diff \alpha \, \e^{-V(\alpha)/\Boltz T}.
    \label{eq:Zcl}
\end{eqnarray}

In the quantum case, the differences between linear and rotational motion are more fundamental. The orientation state of the quantum planar rotor is described by $2\pi$-periodic wave functions $\psi(\alpha) = \la \alpha|\psi\ra$, and the periodicity implies that the angular momentum be quantized in integer multiples of $\hbar$. We can thus express the angular momentum operator as $\oL_z = \sum_{m \in \mathbb{Z}} \hbar m \ket{m} \bra{m}$, defining the orthonormal basis of discrete angular momentum eigenstates, $\la \alpha|m\ra = \e^{im\alpha}/\sqrt{2\pi}$. The expansion coefficients of the wave function in this basis are obtained from its Fourier decomposition.

The conjugate angle operator $\oalpha$ can be defined through the unitary momentum displacement operators $\exp(i k \oalpha)$, which for $k\in\mathbb{Z}$ adhere to the strict periodicity of the system and act like $\exp \left( i k \oalpha \right) \ket{m} = \ket{m+k}$. Consistently, the basis of angle states $|\alpha\ra$ is obtained as
\begin{equation}
    |\alpha\ra = \sum_{m\in \mathbb{Z}} \frac{\e^{-im\alpha}}{\sqrt{2\pi}} |m\ra .
\end{equation}
They form a continuous orthonormal basis, $ \la \alpha_1|\alpha_2\ra = \delta(\alpha_1 - \alpha_2 \mod 2\pi)$, and they are eigenstates of the displacement operators and thus of any periodic function of $\oalpha$ by virtue of the Fourier expansion; e.g., $V(\oalpha)|\alpha\ra = V(\alpha)|\alpha\ra$. The canonical commutation relation between $\oL_z$ and $\alpha$ can be expressed in terms of periodic functions as
\begin{equation}
     \left[ f(\oalpha), \oL_z \right] = i \hbar f^\prime (\oalpha), \quad \forall \,\, 2\pi\text{-periodic} \ f(\alpha) .
    %\left[ \e^{ik\oalpha},\oL_z \right] = -\hbar k \e^{ik\oalpha} \quad \forall \, k\in\mathbb{Z}.
\end{equation}
The quantum version $\oH$ of the generic Hamiltonian \eqref{eq:HamCl} will have a discrete spectrum of energy eigenvalues $E_n$. The corresponding Gibbs state is given by the density matrix $\ogamma = \exp (-\oH/\Boltz T)/Z$ with the quantum partition function $Z = \tr \{ \exp (-\oH/\Boltz T) \}$. 

For our following case studies and numerical computations, we conveniently introduce the rotational energy quantum as a natural energy scale,
\begin{equation}
    E = \frac{\hbar^2}{I},
\end{equation}
and express all relevant energies in units of this scale. When comparing the quantum and the classical rotor as a working medium, we expect notable differences only for thermal energies that do not exceed this scale by far.

\subsection{Otto cycle} 
\label{sec_Otto}
%fix H notation. For classical an index would be better

The Otto cycle is the most widely studied and instructive thermal machine model \cite{Kosloff2017,Feldmann2018}, which can be generically formulated in a classical or quantum setting. We start with the quantum version and introduce the classical counterpart later. The basic setting comprises one hot and one cold thermal reservoir with respective temperatures $T_h > T_c$, and a quantum system acting as the working medium, whose Hamiltonian $\oH(\lambda)$ depends on a control parameter $\lambda$. This parameter can be varied between two extreme values $\lambda_h$ and $\lambda_c$, which abstracts the cyclic motion of a piston. The temperature $T$ and the control parameter $\lambda$ are the two relevant independent state variables of the working medium.

\begin{figure}
    \centering
 \includegraphics[width=0.3\textwidth]{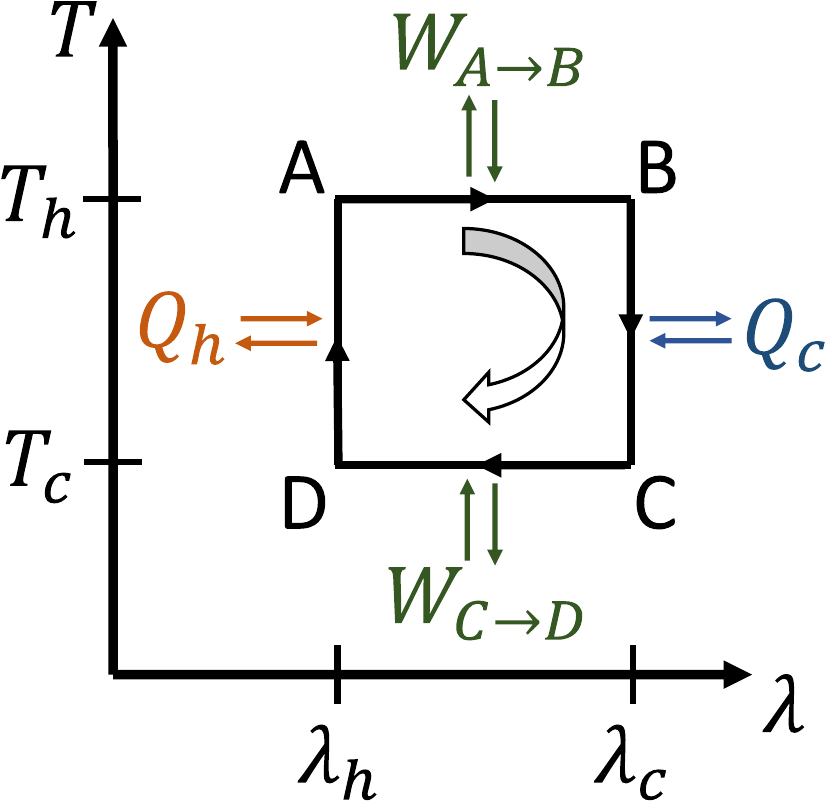}
 \caption{Phase diagram of the four-stroke Otto cycle with a generic working medium in terms of the temperature $T$ and the control parameter $\lambda$. The two horizontal lines represent the isentropic strokes in which the control parameter changes between $\lambda_h$ and $\lambda_c$ under work exchange. The two vertical lines represent ``isochoric'' thermalization of the working medium through heat exchange with a hot reservoir at temperature $T_h$ (left) and a cold reservoir at $T_c$ (right).}
    \label{fig_Otto}
\end{figure}

In its ideal implementation, the Otto cycle consists of four discrete strokes, as sketched in the phase diagram spanned by $\lambda$ and $T$ in Fig.~\ref{fig_Otto}: Two thermalization strokes ($B\to C$ and $D\to A$) in which the system is coupled to either reservoir of temperature $T_h$ or $T_c$ at a respectively fixed control parameter value $\lambda_h$ or $\lambda_c$, and two work strokes ($A\to B$ and $C \to D$) in which the control parameter alternates between $\lambda_h$ and $\lambda_c$ while the system is in isolation. 
Ideally, we assume that the system can fully thermalize with each reservoir, resulting in the two respective Gibbs states
\begin{equation}\label{eq:Gibbs_hc}
    \ogamma_{j} = \frac{1}{Z (\lambda_{j},T_{j})}\exp \left[ -\frac{\oH(\lambda_{j})}{\Boltz T_{j}} \right], \qquad j \in \{h,c\}.
\end{equation}
During the work strokes, we assume that the control parameter is quenched quasi-instantaneously between its two boundary values, so that the thermal populations of the system's energy levels are unaffected and the system remains in one of the two Gibbs states \footnote{In the magnetic dipole setting, an adiabatically slow ramp between $\lambda_h$ and $\lambda_c$ is equivalent to an instantaneous quench, because the energy basis remains unchanged.}. 

Hence, we can identify the net heat input from the hot and cold reservoir as the mean energy change in each respective thermalization stroke, during which the thermal populations change between $\ogamma_h$ and $\ogamma_c$ at a fixed control parameter value,
\begin{align}
    Q_c &= \la H_c \ra_c - \la H_c \ra_h = \tr\{\oH(\lambda_c) (\ogamma_c - \ogamma_h) \} \nonumber \\ 
    Q_h &= \la H_h \ra_h - \la H_h \ra_c = \tr\{\oH(\lambda_h) (\ogamma_h - \ogamma_c) \} .\label{eq:Q_hc}
\end{align}
We abbreviate the mean energy at the control parameter value $\lambda_i$ in thermal equilibrium at temperature $T_j$ by $\la H_i\ra_j = \tr \{ \oH(\lambda_i) \ogamma_j \}$. These mean energy values, four in total, fully characterize the performance and operation mode of the ideal Otto cycle.

In the two work strokes, the change of mean energy is due to a change in the control parameter and can thus be identified as work, \mg{
\begin{align}
    W_{A \to B} &= \langle \oH_c  \rangle_h - \langle \oH_h  \rangle_h
    = \tr \left\{ \left[ \oH(\lambda_c) - \oH(\lambda_h)  \right] \Gibbs_h \right\} \nonumber \\ 
     W_{C \to D} &= \langle \oH_c  \rangle_h - \langle \oH_h  \rangle_h
    = \tr \left\{ \left[ \oH(\lambda_h) - \oH(\lambda_c ) \right] \Gibbs_c \right\}.
\end{align}
} 
Energy conservation over the whole cycle requires that the net total work input of both strokes be
\begin{equation}\label{eq:Wtot}
    W = W_{A \to B} + W_{C \to D} = -(Q_c + Q_h).
\end{equation} 
We distinguish three modes of operation of the so defined ideal Otto cycle. When it yields a net work \textit{output}, $W<0$, it operates as an \textbf{engine} with cycle efficiency $\eta = |W|/Q_h$. When heat is drawn from the cold reservoir, $Q_c >0$, the cycle acts as a \textbf{refrigerator}. In any other case, the cycle is considered useless, acting merely as a \textbf{heater}.

In this paper, the working medium is a classical or quantum planar rotor with a Hamiltonian of the form \eqref{eq:HamCl} or the corresponding quantum version $\oH = H(\oalpha,\oL_z)$. In the two following case studies, the control parameter $\lambda$ modulates the strength of the potential $V$ or the boost frequency $\omega$. For the classical analysis, we simply replace the Gibbs states \eqref{eq:Gibbs_hc} and their partition functions by the respective phase-space quantities, as defined in \eqref{eq:classGibbs} and \eqref{eq:Zcl}, and the energy expectation values in \eqref{eq:Q_hc} by the respective phase-space averages.

\mg{
\subsection{Fluctuations of work}
\label{sec_fluctuation}

In macroscopic working media, fluctuations state variables are typically negligible and thermodynamic process quantities such as heat and work are well represented by their mean values. This is no longer the case in the microscopic regime of working media comprised of only few degrees of freedom, which are subject to comparatively strong (quantum) fluctuations. The output of such thermal machines can then vary randomly and widely across instances or over time. 
To assess the fluctuations in the output of a quantum engine, several notions of work statistics have been proposed, all with their specific advantages and disadvantages \cite{baumer2018fluctuating}. Here we employ a definition based on the ``operator of work'' recently reviewed in \cite{PhysRevResearch.6.L022036}, which yields measurable quantum work statistics for isentropic strokes with a consistent classical limit.

%While heat and work can be typically represented by their mean values in the macroscopic regime, the small systems size and the inherent random nature of quantum theory make fluctuations unavoidable in the microscopic world. Unfortunately, there are various definitions of quantum work statistics with different advantages \cite{baumer2018fluctuating}. However, the approach using the "operator of work" has been recently pointed out to be consistent with many physically motivated criteria \cite{PhysRevResearch.6.L022036}. We choose this approach

Consider a parametric Hamiltonian $\oH(\lambda_t)$ and a time dependent parameter $\lambda_t$ that generates the unitary time evolution operator $\oU(t)$. The operator of work between times $0$ and $t$ is defined as 
\begin{equation}
    \oW_{0 \to t} = \oU^\dagger (t) \oH \left( \lambda_t \right) \oU(t) - \oH(\lambda_0).
\end{equation}
Taking the expectation value of this operator with respect to the initial state $\rho(0)$ results in the average work defined for isentropic strokes as the difference in mean energy,
\begin{align}
    \left\la W_{0 \to t}  \right\ra &= \tr \{ \rho(0)\oW_{0 \to t} \} \nonumber \\
    &= \tr \left\{ \oH ( \lambda_t ) \hat{\rho}(t)   \right\} - \tr \left\{ \oH ( \lambda_0 ) \hat{\rho}(0) \right\}. %    &= W_{0 \to t}
\end{align}
The variance of this observable accordingly provides a measure for the deviations of work around the mean. 

In our case studies below, the Hamiltonian is of the generic form $\oH(\lambda) = \oH_0 + \lambda \oV$ and the work strokes are modeled as quasi-instantaneous quenches. The operators of work associated to the isentropic strokes from point $A$ to point $B$ and from $C$ to $D$ in Fig.~\ref{fig_Otto} reduce to 
\begin{equation}
\label{eq:operator_Work}
    \oW_{A \to B} = (\lambda_c - \lambda_h) \oV = -\oW_{C \to D}.
\end{equation}
Noting that the statistics of these two work operators are evaluated with respect to different initial states $\ogamma_h$ and $\ogamma_c$ and thus independent from each other, we can express the variance of the total per-cycle work in terms of variances of $\oV$ with respect to $\ogamma_{h,c}$,
\begin{equation}
    \label{eq:variance_general}
    \var \left[ W \right] = (\lambda_c - \lambda_h)^2 \left( \var_h[V] + \var_c[V] \right),
\end{equation}
with $\var_h [V] =   \la V^2 \ra_h - \la V \ra_h^2$; see App.~\ref{app_var_work} for details. 
We will quantify the relative amount of work fluctuations in terms of the ``scaled variance'',
%A quantity that describes the ratio between fluctuations of work and the mean work output is the so-called "scaled variance"
\begin{equation}
\label{eq:scaled_variance_general}
    \frac{\var [W ]}{\la W \ra^2} = \frac{\var_h[V] + \var_c[V]}{\left( \la V \ra_c - \la V \ra_h \right)^2}.
\end{equation}
Equations \eqref{eq:variance_general} and \eqref{eq:scaled_variance_general} also apply to the classical case, in which quantum expectation values with respect to \eqref{eq:Gibbs_hc} are replaced by classical phase space averages with respect to \eqref{eq:classGibbs}.
}

\section{Electric dipole machine}
\label{sec_el}

In our first case study, we investigate the setting depicted in Fig.~\ref{fig_Engines}(a): an electric dipole rotating in the $xy$-plane under the influence of a homogeneous electric field (across a capacitor of controlled voltage) pointing in, say, the $x$-direction. 
The control parameter $\lambda$ determines the field strength, $\vec{E}_\lambda = E_\lambda \ve_x$, while the rotor angle $\alpha$ determines the dipole orientation, $\vec{d} = d ( \ve_x \cos \alpha + \ve_y \sin\alpha )$, which results in the potential energy $V(\alpha) = -\vec{E}_\lambda\cdot \vec{d} = -d E_\lambda \cos\alpha $. Writing the control parameter as the dimensionless potential strength $\lambda \equiv 2 d E_\lambda / E $, we arrive at the Hamiltonian
\begin{equation} \label{eq:Ham_el}
    H(\lambda) = \frac{L_z^2}{2I} + E\lambda \sin^2 \left(\frac{\alpha}{2} \right) = E \left[ \frac{L_z^2}{2\hbar^2} + \lambda \sin^2 \left(\frac{\alpha}{2} \right) \right],
\end{equation}
up to an additive constant. This resembles a mathematical pendulum, which behaves approximately like a harmonic oscillator of frequency $\omega_{\rm eff} = \sqrt{E\lambda/2I}$ in the limit of low excitations and $\lambda \gg 1$. We will therefore focus on the rotor-specific regime $\lambda \sim 1$.

\subsection{Classical description}

Describing the electric dipole as a classical planar rotor, the four characteristic mean energy values $\la H_i \ra_j$ of the Otto cycle can be given analytically. They follow from the partition function \eqref{eq:Zcl} of a Gibbs state with respect to the Hamiltonian \eqref{eq:Ham_el},
\begin{equation}
\label{eq:partition_el_class}
    Z(\lambda,T) = \sqrt{2\pi I \Boltz T} \, 2\pi \e^{-E\lambda/2\Boltz T} I_0 \left( \frac{E\lambda}{2\Boltz T} \right),
\end{equation}
The energy values then become
\begin{align}
  \left\langle H_i \right\rangle_j &= \int \diff L_z \int \diff \alpha \, H(\lambda_i) \frac{\exp[-H(\lambda_j)/\Boltz T_j]}{Z(\lambda_j,T_j)}  \nonumber  \\ 
  &= \Boltz T_j \left[\frac{1}{2} - \frac{\lambda_i}{\lambda_j} \frac{ \partial \ln Z(\lambda_j,T_j)}{\partial \lambda_j} \right] \nonumber \\
  &= \frac{\Boltz T_j}{2} + \frac{E \lambda_i}{2} \left[ 1 - \frac{I_1 (x_j)}{I_0 (x_j)} \right],
\end{align}
with $x_j = E \lambda_j/(2 \Boltz T_j)$ and $I_n$ modified Bessel functions. The heat and work inputs per cycle read as
\begin{align}
    Q_c &= \frac{\Boltz (T_c - T_h)}{2} + \frac{E \lambda_c}{2} \left[ \frac{I_1(x_h)}{I_0(x_h)} - \frac{I_1(x_c)}{I_0(x_c)} \right], \\ 
    Q_h &= \frac{\Boltz (T_h - T_c)}{2} + \frac{E \lambda_h}{2} \left[ \frac{I_1(x_c)}{I_0(x_c)} - \frac{I_1(x_h)}{I_0(x_h)} \right], \\ 
     W &= \frac{E(\lambda_h - \lambda_c)}{2} \left[ \frac{I_1(x_h)}{I_0(x_h)} - \frac{I_1(x_c)}{I_0(x_c)} \right]. \label{eq:W_el_cl}
\end{align}
From these we can already infer the modes of operation. To this end, we note that the function $I_1/I_0$ is strictly monotonously increasing and hence, inequalities between function values also hold between the respective arguments. The engine regime, $W<0$, can be achieved in two ways: Either the first factor in \eqref{eq:W_el_cl} is positive ($\lambda_h>\lambda_c$) and the second one is negative ($x_c > x_h$), or vice versa. The latter however implies $T_c >T_h$ and can thus be excluded. As a result, the engine regime is characterized by 
\begin{equation}\label{eq:engineRegime_el}
    W< 0 \quad \iff \quad  \frac{T_h}{T_c} > \frac{\lambda_h}{\lambda_c} > 1.
\end{equation}
The refrigeration regime can be characterized by an implicit inequality only, 
\begin{align}\label{eq:fridgeRegime_el}
    Q_c > 0 \quad \iff \quad \frac{I_1(x_h)}{I_0(x_h)} - \frac{I_1(x_c)}{I_0(x_c)}  > \frac{\Boltz (T_h - T_c)}{E\lambda_c}.
\end{align}
We will now evaluate the classical performance of the engine and refrigerator in terms of the work output and the cold reservoir heat input, respectively, and compare them to the quantum case.

\subsection{Quantum-classical comparison}

For the quantum version of the machine, the characteristic mean energies $\la H_i \ra_j$ can no longer be given analytically. We compute them with help of the quantum optics package for the Julia programming language \cite{Kr_mer_2018}.

\begin{figure}
    \centering
    \includegraphics[width=0.5\textwidth]{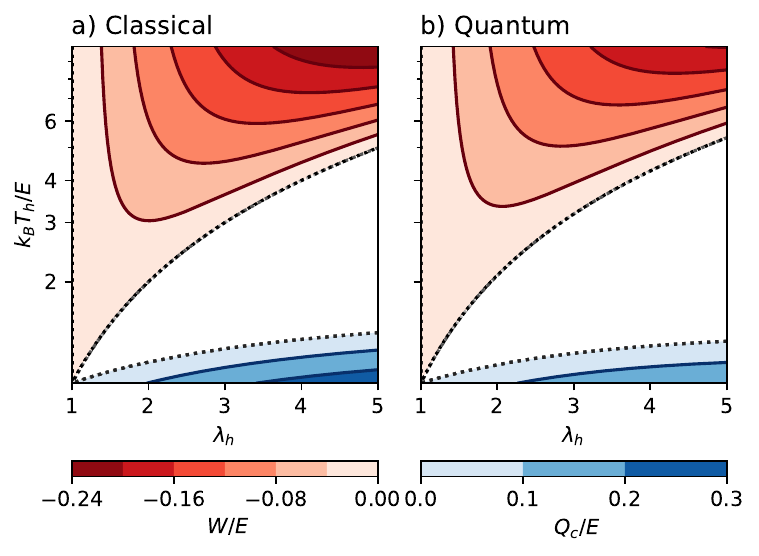}
    \caption{Energy output of an Otto cycle for the electric dipole  machine setting with (a) a classical and (b) a quantum planar rotor. We plot against the hot-stroke control parameter $\lambda_h$ and temperature $T_h$, distinguishing between two operation modes marked by the dotted lines: Engine operation with work output $W<0$ (red shades) and refrigeration with heat output $Q_c>0$ (blue). The cold-stroke parameters are fixed at $\Boltz T_c = E$ and $\lambda_c=1$.}
    \label{fig_El_hot}
\end{figure}

We compare the electric dipole machine output in the classical and the quantum case for an exemplary parameter setting in Fig.~\ref{fig_El_hot}(a) and (b), respectively, against varying hot-stroke field strength and temperature, $\lambda_h$ and $T_h$. The cold-stroke parameters are set to moderate values, $\lambda_c=1$ and $\Boltz T_c = E$. The red- and the blue-shaded countours correspond to different per-cycle outputs of work ($-W$) and cold-bath heat ($Q_c$), respectively, and the dashed lines mark the engine and refrigerator operation regimes. As expected from the classical condition \eqref{eq:engineRegime_el}, we observe that work production occurs when the temperature \mg{ratio is greater than the ratio of dipole potential strength parameters.}
%difference is large and the difference in dipole potential strengths is \mg{comparably} low. 
\mg{The more these parameters (and hence the temperatures) differ, the greater the absolute work output.} Refrigeration is most pronounced in the regime of strongly confined pendulum motion for the hot stroke, $\lambda_h \gg 1$. The quantum rotor behaves similarly to the classical one, exhibiting only a small decrease of its operation regimes and outputs.

\begin{figure}
    \centering
    \includegraphics[width=0.5\textwidth]{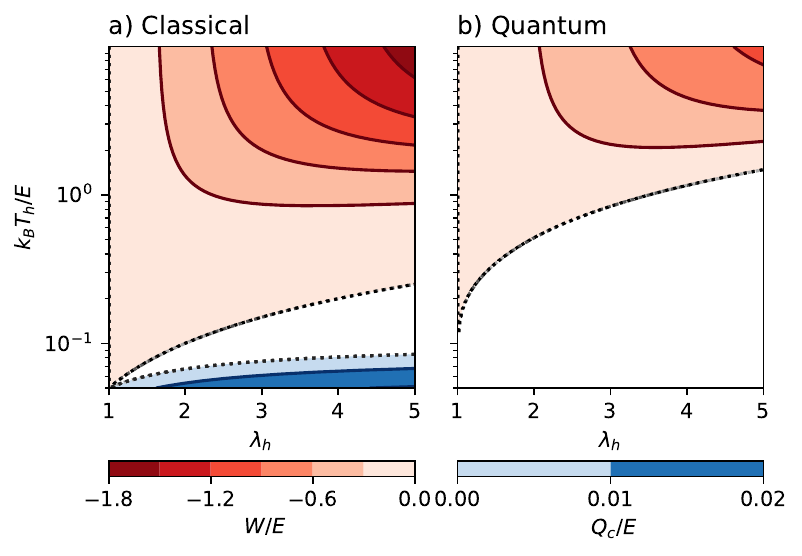}
    \caption{Energy output of (a) the classical and (b) the quantum electric dipole machine, plotted in the same manner as in Fig.~\ref{fig_El_hot} as a function of the hot-stroke control parameter $\lambda_h$ and temperature $T_h$. Here, we set the cold-stroke temperature to a lower value, $\Boltz T_c = 0.05 E$.}
    \label{fig_El_cold}
\end{figure}

More significant differences are observed in Fig.~\ref{fig_El_cold}, where we set the cold bath temperature to $\Boltz T_c = 0.05 E$, close to the ground state. The refrigeration mode is now no longer visible in the quantum case shown in Fig.~\ref{fig_El_cold} (b). Once again, the quantum case is systematically worse in terms of operation regime and output. This agrees with the intuition that quantum Gibbs states of localized or trapped motion typically occupy more phase-space area than classical Gibbs states of the same temperature, and the discrepancy grows at lower temperatures. Hence, from a classical perspective, the quantum machine appears to operate at a lower temperature bias and thus more poorly. 

\begin{figure}
    \centering
    \includegraphics[width=0.5\textwidth]{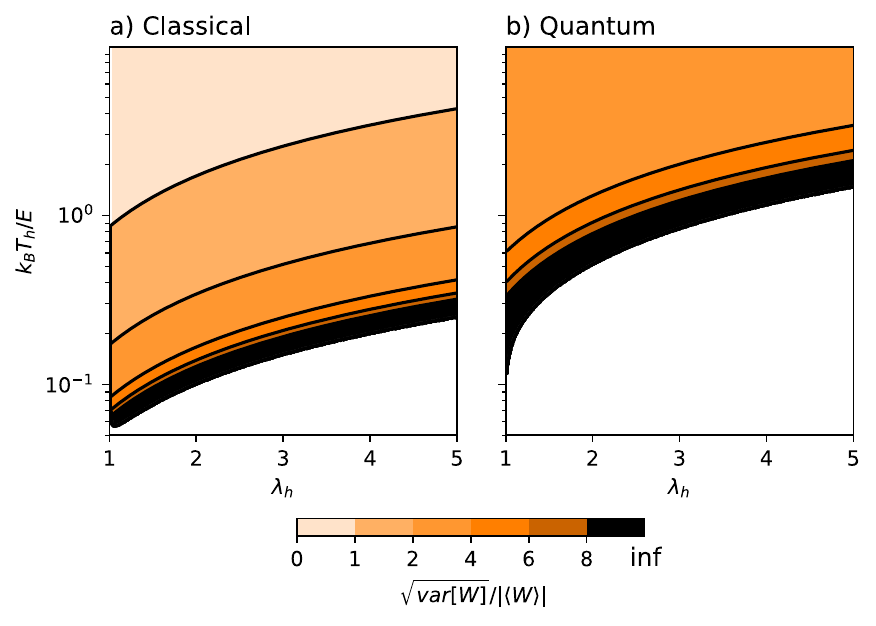}
    \caption{Square root of the scaled work variance for (a) the classical and (b) the quantum electric dipole machine, plotted against against the hot-stroke control parameter $\lambda_h$ and temperature $T_h$ for the same settings as in Fig.~\ref{fig_El_cold}. Unshaded regions correspond to no work output ($\la W \ra \geq 0$), and the black shades represent  a diverging scaled work variance.}
    \label{fig_el_scaled_cold}
\end{figure}

\mg{This intuition is corroborated by the fact that the work fluctuations are consistently higher in the quantum case. Inserting the control Hamiltonian $\oV = E \sin^2 (\hat\alpha/2)$ into \eqref{eq:scaled_variance_general}, we obtain the scaled work variance,
\begin{equation}
\label{eq:scaled_el}
    \frac{\var[W]}{\la W \ra^2} = \frac{\var_h[ \sin^2 \left( \alpha/2 \right)] +\var_c[ \sin^2 \left( \alpha/2 \right)] }{ \left( \la \sin^2 \left( \alpha/2 \right) \ra_h - \la \sin^2 \left( \alpha/2 \right) \ra_c   \right)^2.}
\end{equation}
As shown in App.~\ref{app_var_work}, the classical version of this expression can be given explicitly, whereas the quantum version must be computed numerically. Both versions are compared in Fig.~\ref{fig_el_scaled_cold}, which shows the square root of the scaled variance as a function of $\lambda_h$ and $T_h$ for the same settings as in Fig.~\ref{fig_El_cold}. Not only does the quantum version (b) exhibit greater relative work fluctuations than the classical version (a), but the quantum work variance is also greater than the mean value in this parameter regime. 
%greater in We see that the quantum version possesses consistently higher fluctuations than its classical counterpart. A regime, where $\sqrt{\var[W]} \leq \la W \ra $, can only be observed classically. Quantum-mechanically, the work is always dominated by fluctuations.
}

%However, in the following case study, we will see that this heuristic argument \mg{from before} can fail 
\mg{In the next case study, we will see that a quantum rotor can be beneficial} if the rotational motion is both unrestricted and close to its ground state, as angular momentum quantization then plays a more intricate role.

\section{Magnetic dipole machine}
\label{sec_mag}

We now consider the setting of Fig.~\ref{fig_Engines}(b): a charged dumbbell or rod rotates in the $xy$-plane in the presence of a homogeneous magnetic field perpendicular to that plane, $\vec{B}_\lambda = B_\lambda \ve_z$. Once again, the control parameter $\lambda$ determines the field strength. The charged rotor constitutes a circular current to which one can associate a magnetic dipole moment $\vec{\mu} = \mu L_z \ve_z /\hbar$. Its potential energy in the field, $V = -\vec{\mu}\cdot \vec{B}_\lambda = -\omega_\lambda L_z $, can be given in terms of the controlled Larmor frequency $\omega_\lambda = \mu B_\lambda/\hbar \equiv \lambda \hbar/I$.

The Hamiltonian of the so defined working medium is
\begin{equation}\label{eq:Ham_cl_mag}
    H(\lambda) = \frac{L_z^2}{2I} - \omega_\lambda L_z = E \left[ \frac{L_z^2}{2\hbar^2} - \lambda \frac{L_z}{\hbar} \right],
\end{equation}
where $\hbar\lambda$ determines the net angular momentum displacement that minimizes the energy in the field.
\mg{In both the quantum and the classical case, the relevant mean energies, the total per-cycle work \eqref{eq:Wtot}, and the cold-bath heat input $Q_c$ in \eqref{eq:Q_hc} can then be expressed as
%From \eqref{eq:Ham_cl_mag}, we can derive general expressions for the relevant mean energies with respect to the hot and cold thermal states as well as for work and heat, which hold in both the classical and the quantum case,  
\begin{eqnarray}
    \label{eq:meanEnergy_mag_both}
    \la H_i \ra_j &=& \frac{\la L_z^2\ra_j}{2I} - \frac{E\lambda_i}{\hbar} \la L_z\ra_j, \\
\label{eq:W_mag_both}
    W   &=& E(\lambda_h-\lambda_c) \frac{\la L_z\ra_h - \la L_z\ra_c}{\hbar}, \\
\label{eq:Qc_mag_both}
    Q_c &=& \frac{\la L_z^2\ra_c - \la L_z^2\ra_h}{2I} - E\lambda_c \frac{\la L_z\ra_c - \la L_z\ra_h}{\hbar}.
\end{eqnarray}
}

\subsection{Classical no-go result}
\label{subsec_mag_class}

For a classical planar rotor, we will now show that the magnetic dipole configuration can neither operate as an Otto engine nor as a refrigerator, regardless of the chosen temperatures or control parameters. 

The classical partition function of a Gibbs state is straightforwardly obtained after completing the square in the Hamiltonian \eqref{eq:Ham_cl_mag}, 
\begin{equation} \label{eq:Zcl_mag}
    Z(\lambda,T) = Z(0,T) \e^{E\lambda^2/2\Boltz T} = 2\pi\hbar \sqrt{\frac{2\pi \Boltz T}{E}} \, \e^{E\lambda^2/2\Boltz T}. % \sqrt{(2\pi)^3 I \Boltz T} \, \e^{E\lambda^2/2\Boltz T}.
\end{equation}
From this follow the \mg{ classical values for the mean angular momentum,
\begin{equation}
\langle L_z \rangle_j = \hbar \frac{\Boltz T_j}{E} \frac{\partial}{\partial \lambda_j} \ln Z(\lambda_j,T_j) =  \hbar \lambda_j, \label{eq_L_class_exp}
\end{equation}
and for the relevant mean energies \eqref{eq:meanEnergy_mag_both},
\begin{align}
    \la H_i \ra_j &=  \Boltz T_j \left[ T_j \frac{\partial}{\partial T_j} + (\lambda_j - \lambda_i) \frac{\partial}{\partial \lambda_j} \right] \ln Z(\lambda_j, T_j) \nonumber \\
    %&= \frac{\pi \hbar E}{Z(0,T_j)} \int \diff \ell \, (\ell + \lambda_j)(\ell +\lambda_j - 2\lambda_i) \e^{-E\ell^2/2\Boltz T_j}\nonumber \\
    &= \frac{\Boltz T_j}{2} + \frac{E\lambda_j}{2} (\lambda_j-2\lambda_i). \label{eq:meanEnergy_mag} 
\end{align}
}%
The resulting work \eqref{eq:W_mag_both} and cold-bath heat  \eqref{eq:Qc_mag_both} are
\mg{
\begin{align}
    W   &= E (\lambda_h - \lambda_c)^2 > 0, \label{eq:W_mag}\\ 
    Q_c &= - \frac{\Boltz (T_h-T_c)}{2} - \frac{E}{2}(\lambda_h-\lambda_c)^2 < 0 , \label{eq:Qc_mag}
\end{align}}
which precludes any useful operation of the Otto cycle.

\subsection{Quantum machine operation}
\label{subsec_mag_quant}

\begin{figure}
    \centering
 \includegraphics[width=0.48\textwidth]{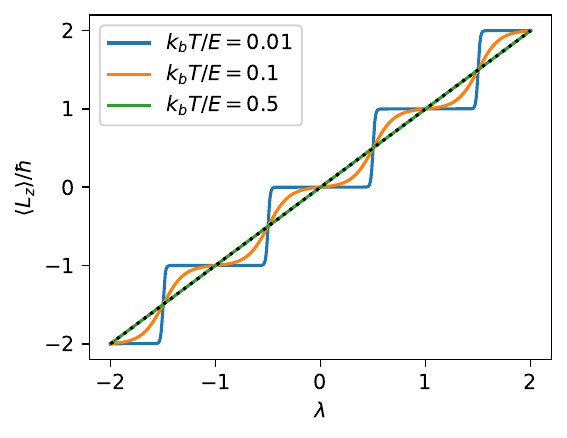}
 \caption{Mean angular momentum of a quantum planar rotor in Gibbs states of three different temperatures, with respect to the magnetic dipole Hamiltonian \eqref{eq:Ham_cl_mag} of varying momentum displacement $\lambda$. The dotted line marks the classical mean value $\la L_z\ra = \hbar\lambda$, achieved at high temperatures.}
    \label{fig_thermal_momentum}
\end{figure}

We will now see that the quantum magnetic dipole setting allows for both an Otto engine and a refrigerator mode, provided that the cold strokes are operated close to the ground state. As before, the performance is determined by the characteristic mean energies $\la H_i\ra_j$, which can be given by derivatives of the partition function as in the second line of \eqref{eq:meanEnergy_mag}. The first lines in \eqref{eq:W_mag} and in \eqref{eq:Qc_mag} also hold, but with the expectation values taken over the quantum Gibbs state at $T_h$ and $T_c$. 
The associated partition function is a discrete sum due to angular momentum quantization. With help of the Poisson sum rule \cite{Pinsky2023}, we can express it as the product of the classical partition function \eqref{eq:Zcl_mag} (in units of Planck's action quantum) and a Jacobi theta function \cite{Whittaker_Watson_2021},
\begin{align}
    Z(\lambda,T) &= \sum_{m=-\infty}^{+\infty} \e^{-E m(m-2\lambda)/2\Boltz T} \nonumber \\ 
    &=
    \sqrt{\frac{2\pi \Boltz T}{E}} \, \e^{E \lambda^2/2\Boltz T} \sum_{\nu= - \infty}^{+ \infty}\!  \e^{ -2\pi \nu (i \lambda + \pi \nu \Boltz T/E)}  \nonumber \\ 
    &= 
    \sqrt{\frac{2\pi \Boltz T}{E}} \, \e^{E \lambda^2/2\Boltz T} \, \vartheta_3 \left( -\pi\lambda, \e^{-2\pi^2 \Boltz T/E} \right).
\end{align}
The quantum expectation value of angular momentum, $\la L_z\ra=: \hbar(\lambda + \epsilon)$, can now deviate from the classical mean value $\hbar\lambda$ 
%\mg{(compare \eqref{eq_L_class_exp})} 
by at most $|\epsilon| < 1/2$. In fact, we can invoke a functional identity of the Jacobi theta function to obtain an explicit Fourier expansion of the deviation \cite{Whittaker_Watson_2021},
\begin{align}
    \epsilon &= \frac{\la L_z\ra}{\hbar} - \lambda = \frac{\Boltz T}{E} \frac{\partial \ln Z(\lambda,T)}{\partial \lambda} -\lambda \nonumber \\
    &= \sum_{n=1}^{\infty} (-)^n \frac{2\pi \Boltz T/E }{\sinh (2\pi^2 n \Boltz T/E)} \sin (2\pi n \lambda). \label{eq:eps_Fourier}
\end{align}
Fig.~\ref{fig_thermal_momentum} plots the resulting $\la L_z\ra$ as a function of $\lambda$ for three temperatures of the underlying quantum Gibbs state. The deviation vanishes exactly for any $T$ whenever $\lambda$ assumes an integer or a half-integer value. Moreover, we observe that the quantum-classical deviation diminishes quickly with growing temperature $T$, and is here no longer visible at $\Boltz T = 0.5E$ (green line). In the opposite limit $\Boltz T \ll E$ (blue curve), the Fourier-sine coefficients in \eqref{eq:eps_Fourier} converge to $(-)^n/\pi n$, which describes a triangular saw-tooth pattern and thus a step-wise increase of 
%\sout{$\la L_z\ra \sout= \hbar \ell$} 
\mg{$\la L_z \ra$} at every half-integer $\lambda$.

To understand this behaviour, recall that the quantum Hamiltonian $\oH(\lambda)$ is diagonal in the angular momentum basis $\{|m\ra \}$. Its energy eigenvalues, $E_m = E(m-\lambda)^2/2 - E\lambda^2/2$, are points on a parabola in $m$ that is centered around $\lambda$. For integer $\lambda$, we have one ground state at $m=\lambda$ and degenerate energy doublets at $m = \lambda \pm n$ with $n\in\mathbb{N}$. Assuming that these doublets are uniformly populated in thermal equilibrium (which would require the  environment to induce incoherent transitions between them upon equilibration), we get $\la L_z\ra = \hbar\lambda$ regardless of temperature. Similarly, for half-integer $\lambda$, all energy states including the ground state are doubly degenerate with respect to $m = \lambda \pm (n-1/2)$, so that once again $\la L_z\ra = \hbar \lambda$. 
Deviations can only occur for $\lambda \mod 1/2 \neq 0$. At low temperatures, the rotor then mainly occupies the angular momentum state \mg{$| [\lambda] \ra$} of minimal energy, \mg{where $[\lambda]$ denotes} the integer closest to $\lambda$. Consequently, \mg{$\la L_z\ra \approx \hbar [\lambda] = \hbar (\lambda +\epsilon)$} with deviation  \mg{$|\epsilon| < \lambda \mod 1/2 <1/2$. In summary, we have
\begin{equation}
\label{eq:hotcold_limit}
    \la L_z \ra \approx \begin{cases}
        \hbar [\lambda] &  \text{for} \ T \to 0 \ \text{and} \ \lambda \notin \mathbb{Z} + \tfrac{1}{2}  \\ 
        \hbar \lambda &  \text{for} \ \lambda \in \mathbb{Z} + \tfrac{1}{2} \\
        \hbar \lambda & \text{for} \ T \to \infty .
    \end{cases}
    % \begin{cases}
    %     \hbar \lambda & \text{for} \ \lambda \ \text{half-integer}  \\
    %     \begin{cases}
    %         \hbar [\lambda]  & \text{for} \ T \to 0 \\
    %   \hbar \lambda    &  \text{for} \  T \to \infty
    %     \end{cases}
    %      & \text{else}
    % \end{cases}
    % \left\{ \begin{array}{lr}
    %      \hbar \lambda & \text{for} \ \lambda \ \text{half-integer}  \\
    %      \left\{ \begin{array}{lr}
    %   \hbar [\lambda]  & \text{for} \ T \to 0 \\
    %   \hbar \lambda    &  \text{for} \  T \to \infty
    % \end{array} \right\} &  \text{else}
    % \end{array} \right\}.
\end{equation}
}

Let us now discuss the implications of quantization for the engine operation regime. We shall restrict our view to the case $\lambda_c > \lambda_h$; the other case could be treated analogously. The per-cycle work \mg{in \eqref{eq:W_mag_both}} can be rewritten as
\begin{equation}
    W = E (\lambda_c-\lambda_h)^2 \left( 1 + \frac{\epsilon_c-\epsilon_h}{\lambda_c-\lambda_h} \right),
\end{equation}
which must be negative (i.e., an output) for an engine. 
The fact that $|\epsilon_c \pm \epsilon_h| < 1$ immediately restricts the choice of control parameters to \mg{ $ |\lambda_c - \lambda_h| < 1$}. Moreover, any integer offset of both parameter values, $\lambda_{h,c} \to \lambda_{h,c}+m$, is irrelevant, \mg{which allows us to restrict our view to $\lambda_{c,h} <1$ without loss of generality.}

We have already seen that there is no work output in the classical regime of vanishing $\epsilon_{h,c}$, i.e., when both temperatures are high, $\Boltz T_{h,c} > E$. In the opposite, deep quantum regime of $\Boltz T_{h,c} \ll E$, we can approximate \mg{$W \approx E (\lambda_c-\lambda_h)([\lambda_c]-[\lambda_h])$ by using \eqref{eq:hotcold_limit} in \eqref{eq:W_mag_both}. %\sout{with the closest integers $l_{h,c}$}. 
This expression} is also non-negative  due to the monotonicity of rounding to the closest integer. \mg{Therefore}, appreciable work output only occurs in an intermediate regime of comparably low $T_c$ and comparably high $T_h$. %\sout{Ideally, a cold stroke at vanishing temperature ($\Boltz T_c \ll E$) and $\lambda_c$ close to $1/2$ should be combined with a hot stroke at high temperature ($\Boltz T_h > E$), so that $\epsilon_c \approx - \lambda_c$ and $\epsilon_h \approx 0$.} 
\mg{In this case, \eqref{eq:hotcold_limit} leads to the average work $W \approx E(\lambda_c-\lambda_h)([\lambda_c] - \lambda_h)$, which becomes negative if and only if $\lambda_c>\lambda_h > [\lambda_c]$. This in turn requires $\lambda_{c,h}<1/2$ and hence $[\lambda_c] = 0$, so that}
%This achieves  
\begin{equation}\label{eq:W_mag_opt}
    \frac{W}{E} \approx -\left( \lambda_c - \lambda_h \right) \lambda_h = \left(  \lambda_h - \frac{\lambda_c}{2}\right)^2 - \frac{\lambda_c^2}{4} > -\frac{1}{16}.
\end{equation}
%In this case
\mg{Here}, setting $\lambda_h = \lambda_c/2$ results in the maximum work output of $E\lambda_c^2/4$ per cycle for a given $\lambda_c \in (0,1/2)$. Hence it is optimal to choose $\lambda_c$ as close as possible (but not identical) to $1/2$. The upper bound of $E/16$ work output can only be achieved asymptotically for $\lambda_c \to 1/2$ and $T_c \to 0$ \footnote{In the alternative case $\lambda_c < \lambda_h$, the maximum work output in \eqref{eq:W_mag_opt} is achievable by setting $\lambda_h=(1+\lambda_c)/2$, in the asymptotic limit of $\lambda_c \to 1/2$ from above, vanishing $T_c$ and high $T_h$.}.

\begin{figure}
    \centering
    \includegraphics[width=0.5\textwidth]{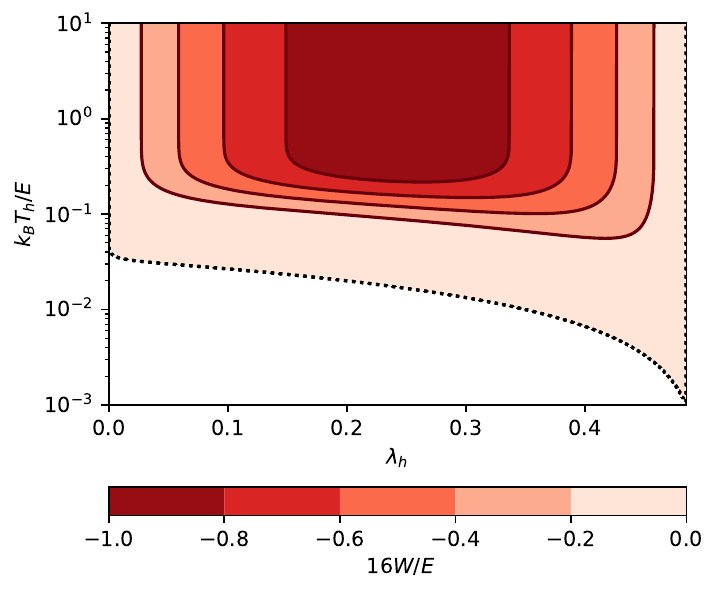}
    \caption{Energy output of an Otto cycle for the magnetic dipole machine setting with a quantum planar rotor. We plot against the hot-stroke control parameter $\lambda_h$ and temperature $T_h$, distinguishing between two operation modes marked by the dotted lines: engine operation with work output $W<0$ (red shades) and refrigeration with heat output $Q_c>0$ (blue). The cold-stroke parameters are fixed at $\Boltz T_c = 0.001 E$ and $\lambda_c = 0.485$. The work output is normalized with respect to theoretically predicted optimum $W = -E/16$, see Equation \eqref{eq:W_mag_opt}.}
    \label{fig_Mag_colder}
\end{figure}

The work output in the ideal engine regime is shown in Fig.~\ref{fig_Mag_colder} as a function of the hot-stroke parameters $\lambda_h$ and $T_h$ for fixed $\lambda_c=0.485$ and $\Boltz T_c = 0.001 E$. We see that the work reaches close to the ideal value around $\lambda_h \approx 0.25$ and for temperatures $\Boltz T_h > 0.2 E$.

\begin{figure}
    \centering
    \includegraphics[width=0.5\textwidth]{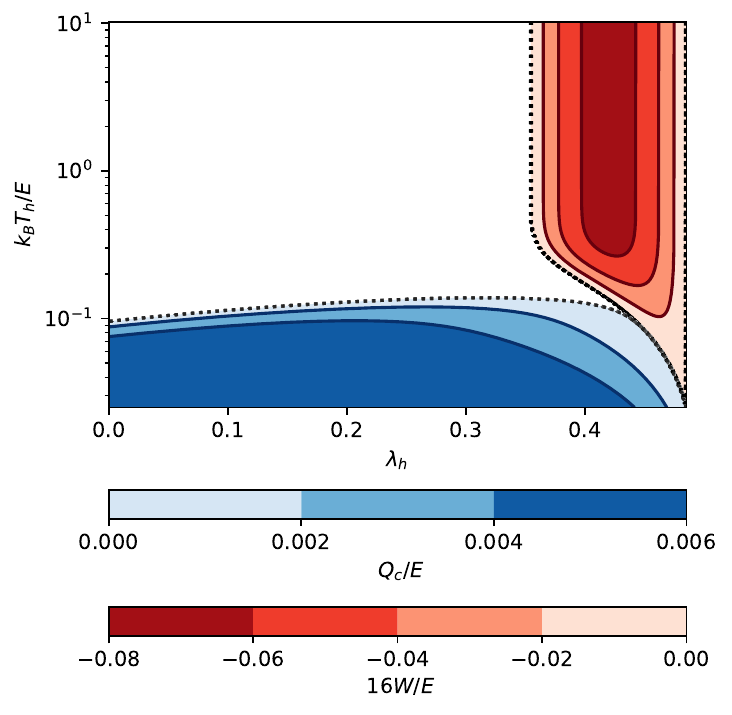}
    \caption{Energy output of the magnetic dipole machine, plotted in the same manner as Fig.~\ref{fig_Mag_colder} against the hot-stroke parameter $\lambda_h$ and the temperature $T_h$. The cold-stroke parameters are fixed at $ \Boltz T_c = 0.025 E$ and $\lambda_c = 0.485$. }
    \label{fig_Mag_cold}
\end{figure}

As one moves away from the ideal regime by increasing the cold temperature, the engine operation window closes and the work output deteriorates quickly. This is illustrated in Fig.~\ref{fig_Mag_cold}, where we set $\Boltz T_c = 0.025 E$. In this setting, we find that the Otto cycle can also operate as a refrigerator, provided that the hot temperature is small, $\Boltz T_h < 0.1 E$. 

\begin{figure}
    \centering
    \includegraphics[width=0.5\textwidth]{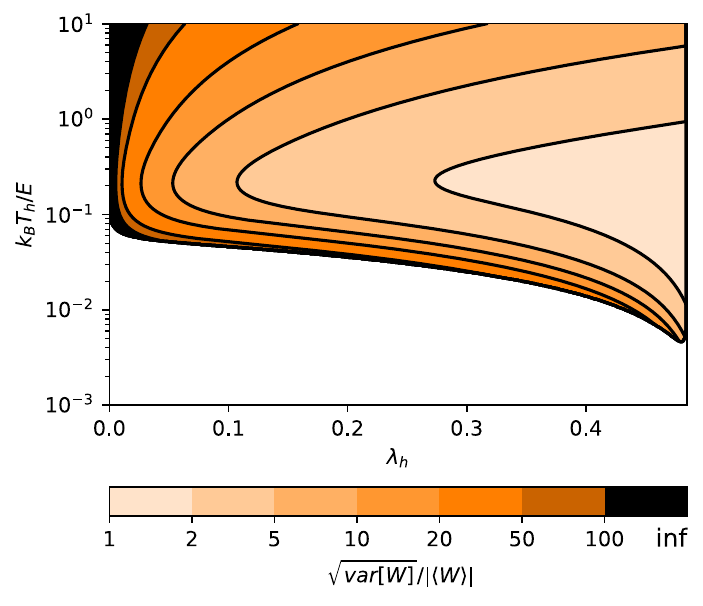}
    \caption{Square root of the scaled work variance for the quantum magnetic dipole machine, plotted against the hot-stroke control parameter $\lambda_h$ and temperature $T_h$ for the same settings as in Fig.~\ref{fig_Mag_colder}. The unshaded region corresponds to no work output, and the black region is where the scaled variance diverges.}
    \label{fig_mag_scaled_colder}
\end{figure}

Overall, we have shown that, because of angular momentum quantization, the magnetic dipole machine supports useful operation modes in the low-temperature regime. 
\mg{However, this regime is dominated by fluctuations, as indicated by the scaled work variance \eqref{eq:scaled_variance_general}. Identifying $\oV = - E \oL_z /\hbar$, it here reads as
\begin{equation}
    \label{eq:scaled_mag}
    \frac{\var \left[ W\right]}{ \la W \ra^2} = \frac{\var_h \left[ L_z \right] + \var_c \left[ L_z \right] }{\left( \la L_z \ra_h - \la L_z \ra_c  \right)^2}.
\end{equation}
This expression can be simplified further in the relevant operation regime of almost zero $T_c$ and high $T_h$, where $\var_c \left[ L_z \right] \approx 0 $, $\la L_z \ra_c \approx 0$, $\la L_z \ra_h \approx \hbar \lambda_h$, and $\var_h[L_z] \approx \hbar^2 \Boltz T_h/E$. The latter is the classical variance calculated in App.~\ref{app_var_work}, and from Fig.~\ref{fig_thermal_momentum}, we infer that $\Boltz T_h \gtrsim 0.5 E$ is already sufficient for the classical limit to be valid. We arrive at
\begin{equation}
    \frac{\var \left[ W\right]}{ \la W \ra^2}  
    \approx \frac{\var_h \left[ L_z \right]}{\hbar^2 \lambda_h^2}  
    = \frac{1}{\lambda_h^2} \frac{\Boltz T_h}{E} \geq 2.
\end{equation}
The lower bound follows, because we must also have $\lambda_c \in (0,1/2)$ for engine operation. The work fluctuations always exceed the mean value, and they grow with increasing hot-bath temperature.
We confirm this observation in Fig.~\ref{fig_mag_scaled_colder}, depicting the square root of the scaled variance \eqref{eq:scaled_mag} for the same parameters as in Fig.~\ref{fig_Mag_colder}.
%This already tells us that the fluctuations dominate in this regime and exceed the mean work. Our rough estimate is supported by the numerical treatment of \eqref{eq:scaled_mag}, that is plotted in Fig.~\ref{fig_mag_scaled_colder}. There we show $\sqrt{\var[W]}/ |\la W \ra|$ against $\lambda_h$ and $T_h$ for fixed $\Boltz T_c = 0.001 E$ and $\lambda_c = 0.485$, which is exactly the parameter space that can be seen in Fig.~\ref{fig_Mag_colder}. It can be seen that the work is indeed dominated by fluctuations since $\sqrt{\var[W]} \geq |\la W \ra |$ in the entire plot.
}

\section{Conclusion}
\label{sec_conclusions}

We investigated two physically motivated realizations of an Otto cycle with a planar rotor as the working medium, highlighting differences in the operation regimes and performance due to quantization. The first realization consists of a rotating electric dipole subject to an electric field of controlled strength parallel to the rotation plane. The system thus resembles a mathematical pendulum. We found that angular momentum quantization leads to consistently lower energy output and smaller engine and fridge operation windows at low temperatures. 

For the second realization, in which a charged rotor generates a magnetic moment subjected to a magnetic field of controlled strength, we showed that there is no useful output in the classical limit. For the quantized rotor, however, we could locate and characterise an engine operation mode in which the rotor state cycles between a deeply quantum cold temperature regime of almost no excitations and a quasi-classical hot regime of arbitrarily high excitations. This constitutes a genuinely quantum thermal machine model that differs from previously studied models based on few-level systems. 

The discussed Otto machine models and their quantum features could be demonstrated in experiments with molecular rotors, levitated nanorotors, or with Josephson loops in circuit QED. 
\mg{For example, the magnetic dipole engine could be demonstrated with a charged nanorotor electrically slowed and aligned in a Paul trap \cite{stickler_quantum_2021}. As a proof pf principle, one could then apply a switchable magnetic field to implement the Otto cycle.}
Future \mg{theoretical} work could explore the possibility of similar quantum features in other mechanical systems with a non-homogeneous energy spectrum.

\acknowledgments
We thank Benjamin A.~Stickler for his valuable suggestions and Tim-Jonas Peter, Marvin Arnold and Florian Sledz for helpful discussions. M.G.~acknowledges support from the House of Young Talents of the University of Siegen.

\appendix

\mg{
\section{Work variance}
\label{app_var_work}

%\subsection{General formula for the variance of work}

In section \ref{sec_fluctuation}, we introduced the work operators 
\begin{equation}
    \oW_{A \to B} = - \oW_{C \to D} = (\lambda_c - \lambda_h) \oV,
\end{equation}
where $\oV$ is the controlled part of the Hamiltonian. The work operators are evaluated with respect to the state of the working medium at the points $A$ and $C$, respectively. For example, the work statistics of the stroke ($A\to B$) is determined by the moments $\tr \{ \gamma_h \oW_{A\to B} \}$.
%at different end points of the Otto cycle. 
The total per-cycle work is the sum of both contributions. 
%In order to keep the following calculations tidy, we make use of the following formal trick: 
Since the end points of the two work strokes are independent, we can conveniently define the total full work operator on the tensor product of the Hilbert space with itself, 
\begin{align}
    \oW &= \oW_{A \to B} \otimes \mathds{1} + \mathds{1} \otimes \oW_{C \to D} \\ 
    &= (\lambda_c - \lambda_h ) \left( \oV \otimes \mathds{1} - \mathds{1} \otimes \oV \right),
\end{align}
and calculate expectation values or higher moments with respect to the state $\ogamma_h \otimes \ogamma_c$. The second moment of work reads as
\begin{equation}
    \la W^2 \ra = (\lambda_c - \lambda_h)^2 \left( \la V^2 \ra_h -2 \la V \ra_h \la V \ra_c +\la V^2 \ra_c \right),
\end{equation}
while the squared expectation value becomes 
\begin{equation}
    \la W \ra^2 = (\lambda_c - \lambda_h)^2 \left( \la V \ra_h^2 -2 \la V \ra_h \la V \ra_c + \la V \ra_c^2   \right).
\end{equation}
Upon subtraction, the mixed term vanishes and we find the work variance
\begin{equation}
    \var \left[ W \right] = (\lambda_c-\lambda_h)^2 \left( \var_h[V] + \var_c[V] \right).
\end{equation}
This expression holds both for the quantum and the classical version.

%\subsection{Electric dipole engine: Variance of (classical) orientation potential}

For the case of the electric dipole engine in section \ref{sec_el}, the controlled part of the Hamiltonian is the pendulum potential, $\oV = E \sin^2 (\hat\alpha/2) $. Its variance with respect to the hot and cold Gibbs state of the classical rotor can be calculated from the first and second moment by taking the first and second derivative of the classical partition function \eqref{eq:partition_el_class},
\begin{align}
    \left\la \left[ E\sin^2 \left(\frac{\alpha}{2}\right) \right]^n \right\ra =\frac{(-\Boltz T)^n}{Z(\lambda,T)} \frac{\partial^n}{\partial \lambda^n } Z(\lambda,T).
\end{align}
Elementary properties of the modified Bessel functions yield
\begin{align}
    \left\la E\sin^2 \left(\frac{\alpha}{2}\right)  \right\ra &= \frac{E}{2} \left(1 - \frac{I_1(x)}{I_0(x)}  \right) \\ 
    \left\la \left[E\sin^2 \left(\frac{\alpha}{2}\right) \right]^2 \right\ra &= \frac{E^2}{2} \left[ 1 - \left(1+\frac{1}{2x} \right) \frac{I_1(x)}{I_0(x)} \right],
\end{align}
with the abbreviation $x = E \lambda/(2 \Boltz T)$. This results in the variance 
\begin{equation}
    \var \left[ E\sin^2 \left( \frac{\alpha}{2}\right) \right] = \frac{E^2}{4} \left[ 1 - \left( \frac{1}{x} + \frac{I_1(x)}{I_0(x)} \right) \frac{I_1(x)}{I_0(x)} \right].
\end{equation}
The quantum counterpart is evaluated numerically.
 
%\subsection{Magnetic dipole engine: Variance of the (classical) angular momentum}
For the magnetic dipole engine in section \ref{sec_mag}, which operates only in the quantum regime, the control Hamiltonian is $\oV = E \lambda \oL_z/\hbar $. Rewriting Eq.~\eqref{eq:meanEnergy_mag_both} and using that the hot thermal state yields approximately the classical expectation value $\la L_z \ra_h \approx \hbar \lambda_h$ from \eqref{eq_L_class_exp}, we obtain
\begin{align}
    \la  L_z^2 \ra_h &= 2 \hbar^2 \left( \frac{\la H_h \ra_h}{E} + \lambda \frac{\la L_z \ra_h}{\hbar} \right) \nonumber \\
    &\approx \frac{2 \hbar^2}{E} \la H_h \ra_h + 2 (\hbar \lambda_h)^2,
\end{align}
Hence, the angular momentum variance is directly given by the expectation value of $\oH_h$,
\begin{equation}
    \var_h \left[ L_z \right] = \hbar^2 \left( \frac{2 \la H_h \ra_h}{E} + \lambda_h^2 \right),
\end{equation}
which can be read off Eq.~\eqref{eq:meanEnergy_mag}. We get 
\begin{align}
    \var_h \left[ L_z \right] %&= \hbar^2 \left( \frac{2}{E} (\Boltz T_h/2 - E \lambda_h^2 /2) + \lambda_h^2 \right) \nonumber  \\
    &= \frac{\hbar^2 \Boltz T_h}{E} = I \Boltz T_h.
\end{align}

}

\end{document}